\documentclass{article}

\usepackage{arxiv}
\usepackage[numbers]{natbib}

\usepackage[utf8]{inputenc} 
\usepackage[T1]{fontenc}    
\usepackage{hyperref}       
\usepackage{url}            
\usepackage{booktabs}       
\usepackage{amsfonts}       
\usepackage{nicefrac}       
\usepackage{microtype}      
\usepackage{lipsum}		
\usepackage{graphicx}
\usepackage{natbib}
\usepackage{doi}

\usepackage{bm}
\usepackage{amssymb}
\usepackage{subfigure}
\usepackage{float}
\usepackage{tabularx, booktabs}
\usepackage{bm}
\usepackage{amsmath}
\usepackage{mathtools}
\usepackage{hyperref}
\usepackage[usenames]{color}

\title{Design and comparison of two linear controllers with precompensation gain for the Quadruple inverted pendulum }

\author{ \href{https://orcid.org/0000-0002-1882-6973}{\includegraphics[scale=0.06]{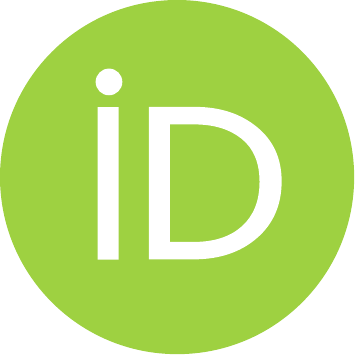}\hspace{1mm}Franklin Josue Ticona Coaquira} \\
	Department of mechatronic engineering\\
	Universidad Católiva Boliviana San Pablo\\
	La Paz, Bolivia \\
	\texttt{franklin.ticona@ucb.edu.bo}
}

\date{}

\renewcommand{\headeright}{Technical Report}
\renewcommand{\undertitle}{Technical Report}
\renewcommand{\shorttitle}{Quadruple inverted pendulum control}

\hypersetup{
pdftitle={A template for the arxiv style},
pdfsubject={q-bio.NC, q-bio.QM},
pdfauthor={David S.~Hippocampus, Elias D.~Striatum},
pdfkeywords={First keyword, Second keyword, More},
}

\begin{document}
\maketitle

\begin{abstract}
In this work we present a workflow for designing two linear control techniques applied to the dynamic system quadruple inverted pendulum mounted on a cart (QIP) where the steady state error on cart position is eliminated through a precompensation gain. The first control law designed was based on LQR, technique that stabilizes the system states based on a minimization of the total energy of the system states and the second control technique was pole placement by 2nd order system approximation, this method allowed to define a settling time and overshoot of the desired response of the plant, for this control the system was considered SISO, being the only output of the plant the horizontal position of the cart. To eliminate the steady state error generated by both linear controllers, an analysis of the stationary response of the system under control in Laplace Domain was performed. The simulation environment used to validate the results presented in this paper was Simscape of Matlab, it is a simulation environment that allowed to simulate the dynamic of nonlinear mechanical systems. The results obtained indicate that both controllers are able to stabilize the plant and move the cart position to the desired setpoint. Finally, we discuss about the performances of both linear control techniques. 
\end{abstract}


\section{Introduction}
\begin{figure}[H]
	\centering
	\subfigure[Schematic diagram of QIP]{\includegraphics[width=0.48\linewidth]{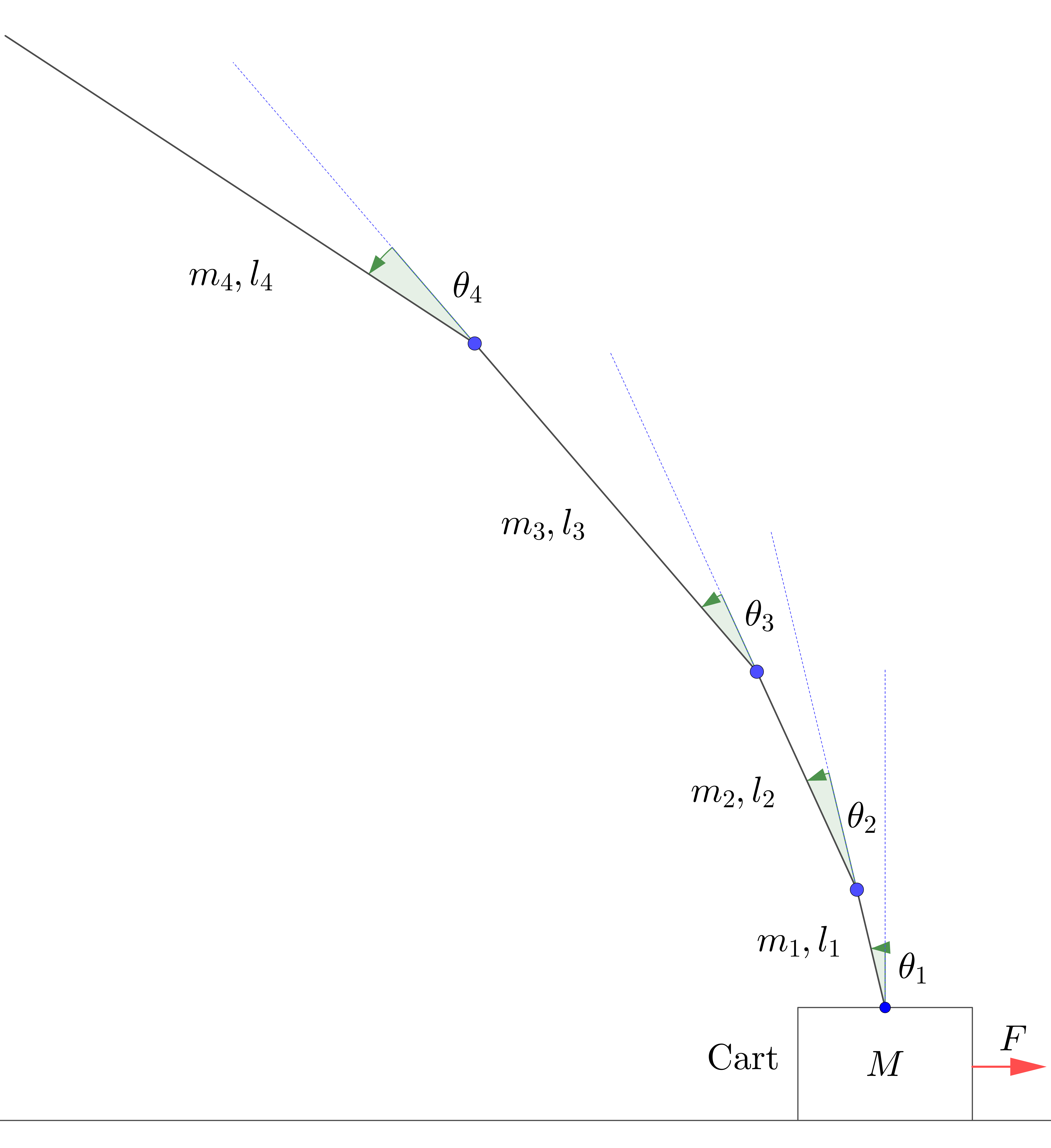}}
	\subfigure[Reference frames]{\includegraphics[width=0.48\linewidth]{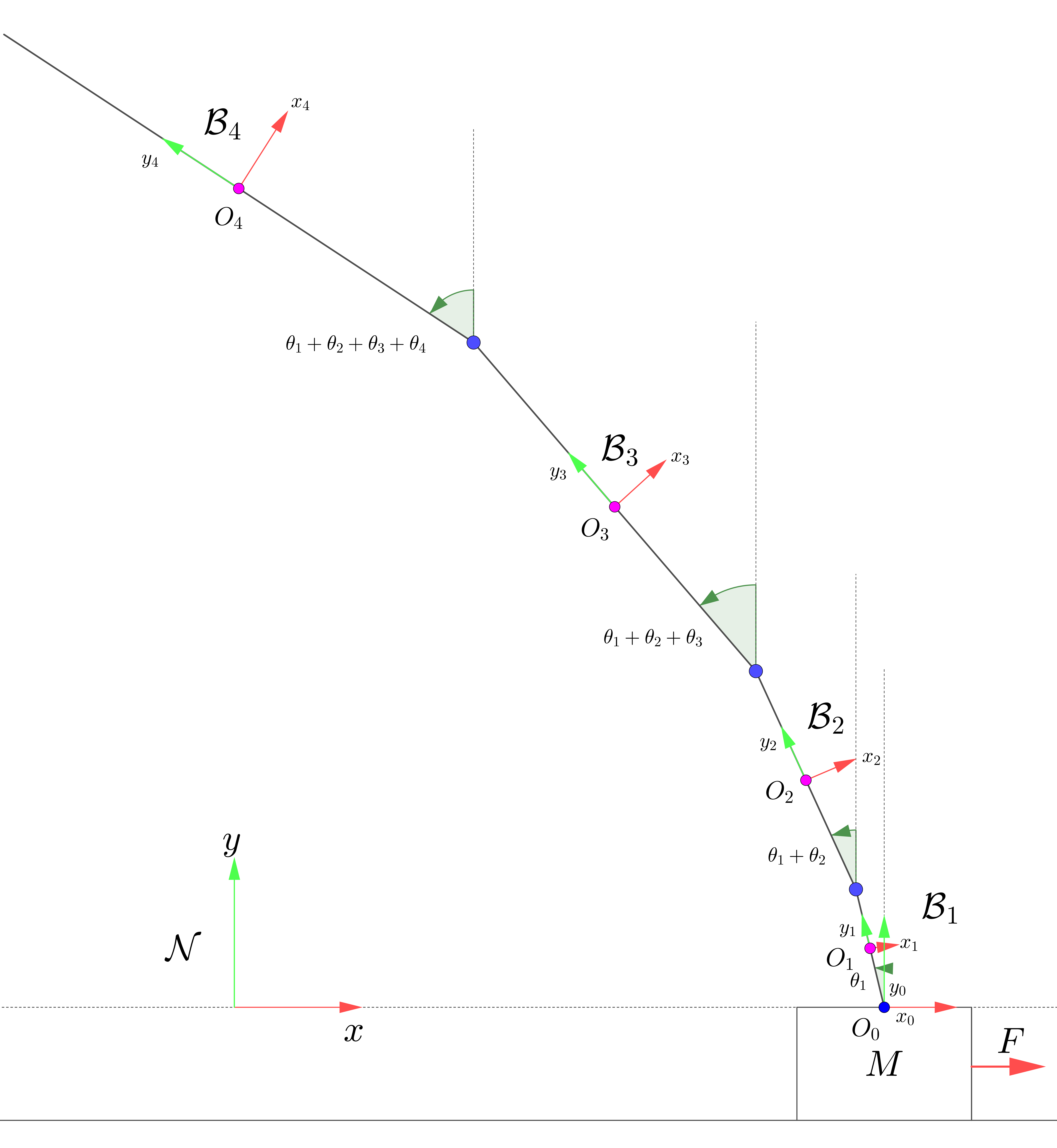}}
	\caption{Kinematic diagram of QIP}
	\label{Kinematic_diagram}
\end{figure}

The inverted pendulum dynamic system is a nonlinear plant consisting of two rigid bodies, the first a block with prismatic motion and the second a link connected to the block by a revolute joint that is usually modeled as a bar, to modify the dynamics of this system an input force is applied directly to the cart. This system turns out to be one of the most common when testing the stability and efficiency of various control laws \cite{bradshaw_shao_1996}. Apart from being able to analyze the performance of various control laws, there are other variants of this system that allow the dynamic behavior of other systems to be modeled, for example \cite{zurawska_reconfigurable_2017} models the motion of a biped robot using an inverted double pendulum, \cite{zhen_inverted_2018} models lateral pedestrian-walkway interaction behavior using an inverted pendulum mounted on a cart. The autor \cite{collini_oscillations_2016} models the vibrations of a construction by means of a simple inverted pendulum. Given the versatility of modeling various dynamic systems, this paper proposes to model and control the inverted quadruple pendulum plant applying the linear control techniques LQR and pole placement, in the latter case we can design the control law considering a settling time and overshoot of the system under control. Furthermore, the steady state error in both cases is eliminated.\\
In this paper the following mathematical notation is considered to model QIP. State vector:
\begin{equation}
   \bm{x}=
    \begin{bmatrix}
           x &
           \dot{x}&
           \theta_{1}&
           \dot{\theta}_1 &
           \theta_{2}&
           \dot{\theta}_2 &
           \theta_{3}&
           \dot{\theta}_3 &
           \theta_{4}&
           \dot{\theta}_4
         \end{bmatrix}^T
\label{state_variables}
\end{equation}
Where $\theta_i$ represents the angle of the link $i$ and $x$ represents the cart's position. We consider that $\theta_i$ ($i\in\{1,2,3,4\}$) is positive in the counterclockwise direction of the clock, considered zero if link $i$ is parallel to link $i-1$. (for $i=1$, $\theta_1=0$ if the first link is parallel to the axis of the ordinates of the inertial reference frame), $x$ is measured from the inertial reference frame, all of these elements can be seen in figure \ref{Kinematic_diagram}.

\section{Related works}
The design of linear controllers for higher order plants is widely studied in the literature, for example \cite{higher_order} shows that with the pole placement technique it is possible to control a conveyor belt (5th order) and an aerial vehicle (6th order). In order to facilitate the design of high order control systems based on pole placement \cite{PID_pole} studied an approximation method for a second order system for the implementation of PID controllers. Regarding the stability and controllability of inverted pendulums \cite{8619302} showed that an $n$-th inverted pendulum is stable in all its states and controllable with respect to the position of the cart. The work \cite{7976186} makes a study about the identification of the states of an inverted pendulum with $n$-links, showing a robust technique to identify the states of the system. Regarding LQR control, authors like \cite{LQR1} obtain an optimal value of the feedback vector $\bm{K}$ by employing adaptive methods, \cite{LQR2} shows a workflow based on Particle Swarm Optimization to obtain the most efficient values of the weighing matrices $\bm{Q}$ and $\bm{R}$ obtaining efficient setpoint responses. On the other hand, for a more realistic simulation of highly nonlinear electromechanical systems, authors such as \cite{simscape1,simscape2} employ the Simscape simulation environment to simulate biomechanical systems, \cite{simscape3} design a control for a 5 DOF robot arm and contrast their results using Simscape. The autor \cite{simscape4} indicates that Simscape is suitable for simulating highly nonlinear mechatronic systems given its modular simplicity and reliability of results in functional prototyping. Due to all mentioned this work presents the comparison of LQR control and pole placement control by second order approximation system applied to the dynamic system quadruple inverted pendulum, besides their performance will be analyzed with the purpose of finding the most suitable linear controller applied to this plant of 10th order. 
\section{Materials and methods}
In this section we show the mathematical modeling of the inverted quadruple pendulum mounted on a cart, generating a heuristic model for $n$ links. In addition, the feedback vector and precompesation gain, wich eliminates the steady state error, for LQR and pole placement is found. Finally, the block diagram used in Simscape for the nonlinear simulation of the plant is presented.
\subsection{Mathematical modeling}
The purpose of this section is to find the linearized mathematical model of QIP using Lagrangian mechanics and linearization of the plant model through Taylor series approximation.\\
Considering the bindings present in the system: cart bound to move in a line and the 4 links bound to move rotationally with respect to the previous one, we can conclude that QIP has 5 generalized coordinates:
\begin{equation}
   \bm{q}=
    \begin{bmatrix}
           x &
           \theta_{1}&
           \theta_{2}&
           \theta_{3}&
           \theta_{4}
         \end{bmatrix}^T
\end{equation}
Where $\bm{q}$ is the vector of generalized coordinates. According to Lagrangian mechanics, the mathematical modeling of this system can be obtained by:
\begin{equation}
	\frac{d}{dt}\nabla_{\dot{\bm{q}}}L(t,\bm{q},\dot{\bm{q}})-\nabla_{\bm{q}}L(t,\bm{q},\dot{\bm{q}})=\bm{Q}_{no-pot}
	\label{euler_lagrange}
\end{equation}
Where $L(t,\bm{q},\dot{\bm{q}})$ is defined as the Lagrangian of the system. The function $L$ is defined as:
	\begin{equation}
	L(t,\bm{q},\dot{\bm{q}})=T(t,\bm{q},\dot{\bm{q}})-V(t,\bm{q})
	\label{lagrangian}
	\end{equation}
Where $T$, $V$ are the total kinetic and potential energy of the system respectively. We will consider that the cart and links are rigid bodies. The kinetic energy of a rigid body is given by:
\begin{equation}
	T=T_{O_i}+T_{rel,O_i}+M\langle \bm{v}_{O_i},\bm{v}_{cm,O_i}\rangle
\end{equation}
If we consider the pivot $O_i$ is the center of mass of each link, then $M\langle \bm{v}_{O_i},\bm{v}_{cm,O_i}\rangle=M\langle \bm{v}_{O_i},\bm{v}_{O_i,O_i}\rangle=0$, thus the kinetic energy of the link $i$ is:
\begin{equation}
	T_i=T_{cm-i}+T_{rel,cm-i}
\end{equation}
Thus, the total kinetic energy of QIP is:
\begin{equation}
	T=T_M+\sum_{i=1}^{4}\left( T_{cm-i}+T_{rel,cm-i}\right) 
	\label{total_kinetic}
\end{equation}
For the purpose of finding $T$, the cart binding equation is identified:
\begin{equation}
	\bm{r}_{cm-M}=(x,0,0)
\end{equation}
Now let's find the vector equation of the center of mass of each link, remembering that the measurement of $\theta_i$ is made with respect to the previous link. Vector of center of mass of the link $i$:
\begin{equation}
	\bm{r}_{cm-i}=( x-\sum_{j=1}^{i-1}(l_j\sin{\sum_{k=1}^{j}\theta_k})-\frac{l_i}{2}\sin{\sum_{j=1}^{i}\theta_j},\sum_{j=1}^{i-1}(l_j\cos{\sum_{k=1}^{j}\theta_k})+\frac{l_i}{2}\cos{\sum_{j=1}^{i}\theta_j},0) 
\end{equation}
Angular velocity of the link $i$:
\begin{equation}
    \bm{\omega}_{O_i}=(0,0,\sum_{j=1}^{i}\dot{\theta}_j)
\end{equation}
\begin{figure}[t]
	\centering
	\includegraphics[width=260px]{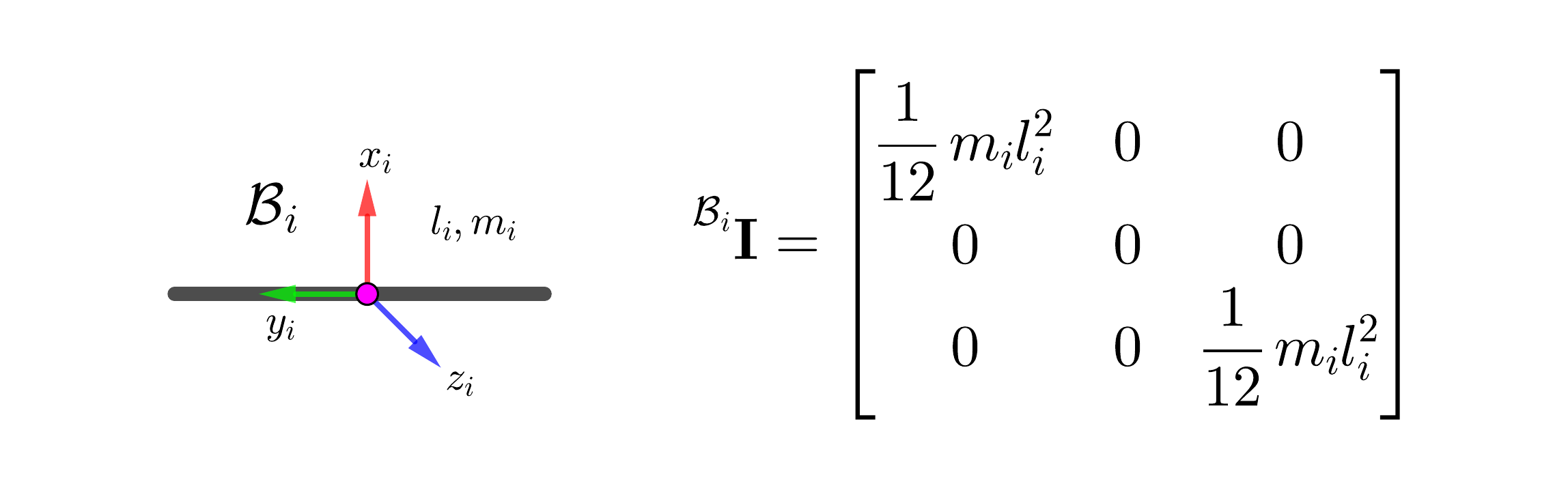}
	\caption{Reference system and inertia tensor of the link $i$}
\end{figure}
Replacing the obtained in (\ref{total_kinetic}), we obtain that:
\begin{equation}
	T=\frac{1}{2}M|\dot{\bm{r}}_{M}|^2+\sum_{i=1}^{4}\left( \frac{1}{2}m_i|\dot{\bm{r}}_{cm_i}|^2+\frac{1}{2}[\omega_{\mathcal{B}_i}]^T\prescript{\mathcal{B}_i}{}{\mathbf{I}}[\omega_{\mathcal{B}_i}]\right) 
	\label{k_kin}
\end{equation}
Furthermore, the potential energy of the link $i$ is given by:
\begin{equation}
	V_i=m_ig\left[ \sum_{j=1}^{i-1}(l_j\cos{\sum_{k=1}^{j}\theta_k})+\frac{l_i}{2}\cos{\sum_{j=1}^{i}\theta_j}\right] 
\end{equation}
Thus, the total potential energy of QIP is given by:
\begin{equation}
	V=\sum_{i=1}^{4}\left\lbrace m_ig\left[ \sum_{j=1}^{i-1}(l_j\cos{\sum_{k=1}^{j}\theta_k})+\frac{l_i}{2}\cos{\sum_{j=1}^{i}\theta_j}\right] \right\rbrace 
	\label{v_pot}
\end{equation}
Finally, we can obtain the Lagrangian of QIP by replacing (\ref{k_kin}) and (\ref{v_pot}) in (\ref{lagrangian}) and thus the mathematical model of QIP will be given by (\ref{euler_lagrange}). Following this recursive process, the mathematical model can be found.\\
Once the dynamic QIP equations are obtained, we can use the compact form of the Euler Lagrange equations:
\begin{equation}
    M(\bm{q})\ddot{\bm{q}}+C(\bm{q},\dot{\bm{q}})\dot{\bm{q}}+D\dot{\bm{q}}+G(\bm{q})=\bm{F}
    \label{e_l_f}
\end{equation}
Usually, this form is employed to analyze and design nonlinear controllers \cite{973983}. In order to obtain a linear state space, we can make use of (\ref{e_l_f}) to linearize the equations of motion of QIP. Applying matrix algebra on (\ref{e_l_f}):
\begin{equation}
	\ddot{\bm{q}}=M^{-1}(\bm{q})[\bm{F}-(C(\bm{q},\dot{\bm{q}})+D)\bm{q}-G(\bm{q})]    
	\label{line}
\end{equation}
Considering the state variables (\ref{state_variables}), we can express the mathematical model in the following form:
\begin{equation}
    \dot{\bm{x}}=f(\bm{x},\bm{F})
    \label{sss_from}
\end{equation}
To proceed with the linearization of the nonlinear model obtained, the linearization points must be found where the plant can be stabilized. Physically, it is known that the points of maximum and minimum potential energy of a system are its equilibrium points. Therefore, the following QIP equilibrium points are chosen: $x_{2-eq}=x_{3-eq}=x_{4-eq}=x_{5-eq}=x_{6-eq}=x_{7-eq}=x_{8-eq}=x_{9-eq}=x_{10-eq}=0$ which in turn will be the operating points for the linearization of the system. Once the operating points of the plant have been obtained, the Taylor series approximation of the nonlinear mathematical model can be applied and taken to a linear state space as follows:
\begin{equation}
    \dot{\bm{x}}=\left[\frac{\partial f_i(\bm{x},\bm{F})}{\partial x_j} \right]  _{\bm{x}_{eq},\bm{F}_{eq}}\bm{x}+\left[ \frac{\partial f_i(\bm{x},\bm{F})}{\partial F}\right] _{\bm{x}_{eq},\bm{F}_{eq}}\bm{F}
\end{equation}
Furthermore, the output vector is:
\begin{equation}
\bm{y}=
\left[ \begin{array}{cccccccccc}
1 & 0 & 0 & 0 & 0 & 0 & 0 & 0 & 0 & 0
\end{array}\right]
\bm{x}
\end{equation}
The physical parameters of QIP considered in this document can be found in the table \ref{table_1}.

\begin{table}[!t]

\caption{Physical parameters of QIP}
\label{table_1}
\centering

\begin{tabularx}{\linewidth}{XXX}
\toprule
Symbol&physical quantity & value \\ \midrule
	$m_1$& link 1, mass [Kg]&$0.1$  \\ \midrule
	$m_2$& link 2, mass [Kg]&$0.1$  \\ \midrule
	$m_3$& link 3, mass [Kg]&$0.1$  \\ \midrule
	$m_4$& link 4, mass [Kg]&$0.1$  \\ \midrule
	$l_1$& link 1, length [m]&$0.03$  \\ \midrule
	$l_2$& link 2, length [m]&$0.04$  \\ \midrule
	$l_3$& link 3, length [m]&$0.07$  \\ \midrule
	$l_4$& link 4, length [m]&$0.10$  \\ \midrule
	$g$& gravity [m/$s^2$]&$9.81$  \\ 
\bottomrule
\end{tabularx}
\end{table}
Thus, the matrices $\bm{A}$, $\bm{B}$, $\bm{C}$ and $\bm{D}$ of the linearized state space of QIP are:
\begin{center}
    $	\bm{A}=
	\begin{bmatrix}
		 0 & 1 & 0 & 0 & 0 & 0 & 0 & 0 & 0 & 0  \\
		0 & 0 & 28.25280 & 0 & -5.53284 & 0 & 0.94176 & 0 & -0.11772 & 0 \\
		0 & 0 & 0 & 1 & 0 & 0 & 0 & 0 & 0 & 0  \\
		0 & 0 & 1608.84000 & 0 & -1659.85000 & 0 & 282.52800 & 0 & -35.31600 & 0 \\
		0 & 0 & 0 & 0 & 0 & 1 & 0 & 0 & 0 & 0  \\
		0 & 0 & -1932.57000 & 0 & 3375.62000 & 0 & -1200.74000 & 0 & 150.09300 & 0\\
		0 & 0 & 0 & 0 & 0 & 0 & 0 & 1 & 0 & 0  \\
		0 & 0 & 374.18100 & 0 & -1983.16000 & 0 & 1634.63000 & 0 & -361.98900 & 0\\
		0 & 0 & 0 & 0 & 0 & 0 & 0 & 0 & 0 & 1  \\
		0 & 0 & -62.22340 & 0 & 329.78400 & 0 & -883.57300 & 0 & 599.19500 & 0 \\
	\end{bmatrix},
\bm{B}=\begin{bmatrix}
	0\\
	7.7600\\
	0\\
	328.0000\\
	0\\
	-394.0000\\
	0\\
	76.2857\\
	0\\
	-12.6857
\end{bmatrix}$
\end{center}
\begin{equation*}
\bm{C}=\left[ \begin{array}{cccccccccc}
1 & 0 & 0 & 0 & 0 & 0 & 0 & 0 & 0 & 0
\end{array}\right], \bm{D}=[0]
\end{equation*}
\subsection{Linear Quadratic Regulator}
Given the dynamic system:
\begin{equation}
	\dot{\bm{x}}(t)=\bm{A}\bm{x}(t)+\bm{B}\bm{u}(t)
\end{equation}
And the cost function defined by:
\begin{equation}
	J=\int_{0}^{\infty}\left( \bm{x}(t)^T\bm{Q}\bm{x}(t)+\bm{u}(t)^T\bm{R}\bm{u}(t)\right) dt
	\label{cost}
\end{equation}
Where $\bm{Q},\bm{R}$ are positive semidefinite matrices. If the control law is defined:
\begin{equation}
	\bm{u}(t)=-\bm{K}\bm{x}(t)
\end{equation}
Then, the matrix $\bm{K}$ that minimizes $J$ and stabilizes the states of the system is:
	\begin{equation}
		\bm{K}=\bm{R}^{-1}\bm{B}^{T}\bm{P}
		\label{const_k}
	\end{equation}
Where $\bm{P}$ is the solution of the algebraic Ricatti equation:
	\begin{equation}
		\bm{P}\bm{A}+\bm{A}^T\bm{P}-\bm{P}\bm{B}\bm{R}^{-1}\bm{B}^{T}\bm{P}+\bm{Q}=\bm{0}
		\label{Ricatti}
	\end{equation}
Therefore, the value of the feedback vector $\bm{K}$ will be found. First, let us define $\bm{Q}=diag(10,1,10,1,1,10,1,1,10,1,1,10,1,10,1)$, we select this value due to we give more energy importance to the cart position and angular positions of the links in the cost function (\ref{cost}), and the matrix $\bm{R}=[1]$, due to too much energy importance at the input would imply that QIP may go out of its linear zone. Solving (\ref{Ricatti}) for $\bm{K}$, we obtain that:
\begin{equation*}
	\scriptsize
	\bm{K}=
	\begin{bmatrix}
		3.16\\3.68\\-14.60\\-5.75\\-163.85\\-5.33\\529.74\\1.78\\-578.51\\-25.21
	\end{bmatrix}
\end{equation*}
Since this vector exists, QIP under this control is stable. However, while the angular positions of the links become stable, we must perform a steady-state stability analysis for the position of the cart. The closed-loop state space of the controlled system is:
\begin{equation}
	\dot{\bm{x}}(t)=(\bm{A}-\bm{B}\bm{K})\bm{x}(t)+\bm{B}\bm{r}(t)
\end{equation}
\begin{equation}
	\bm{y}(t)=\bm{C}\bm{x}(t)
\end{equation}
Considering that this system is SISO, the only output of QIP is considered to be the position of the cart (the angular positions of the links are stabilized at zero), so the following transfer function in the Laplace domain can be obtained:
\begin{equation}
    G_{cl}(s)=\bm{C}(s\bm{I}-\bm{A}+\bm{B}\bm{K})^{-1}\bm{B}
\end{equation}
Considering the error $e(t)=r(t)-y(t)$, in Laplace's domain:
\begin{equation}
    \bm{E}(s)=\bm{R}(s)-\bm{Y}(s)
\end{equation}
Applying the end-value theorem:
\begin{equation*}
    \lim\limits_{t\to\infty}e(t)=\lim\limits_{s\to0}s\bm{E}(s)=\lim\limits_{s\to0}s(\bm{R}(s)-\bm{Y}(s))
\end{equation*}
\begin{equation*}
    \lim\limits_{t\to\infty}e(t)=\lim\limits_{s\to0}s(1-G_{cl}(s))\bm{R}(s)
\end{equation*}
Considering the unit step as the reference input of the system $\bm{R}(s)=N\cdot\frac{1}{s}$ with $N$ a precompensation gain:
\begin{equation}
    \lim\limits_{t\to\infty}e(t)=1+N\bm{C}(\bm{A}-\bm{B}\bm{K})^{-1}\bm{B}
    \label{error}
\end{equation}
Since what is desired is $\lim\limits_{t\to\infty}e(t)=0$, this value is replaced in (\ref{error}). Solving the equation for $N$ we obtain that:
\begin{equation}
    N=[-\bm{C}(\bm{A}-\bm{B}\bm{K})^{-1}\bm{B}]^{-1}
    \label{error2}
\end{equation}
The latter expression being the compensation gain to eliminate the steady state error. Numerically for the LQR control it is $N_{LQR}=3.1623$.
\subsection{Pole placement by 2nd order system approximation}
Given the dynamic system:
\begin{equation}
	\dot{\bm{x}}(t)=\bm{A}\bm{x}(t)+\bm{B}\bm{u}(t)
\end{equation}
\begin{equation}
\bm{y}(t)=\bm{C}\bm{x}(t)
\end{equation}
Its characteristic polynomial is given by:
\begin{equation}
    \det(s\bm{I}-A)
\end{equation}
The pole placement technique defines the law of control.:
\begin{equation*}
    \bm{u}(t)=-\bm{K}\bm{x}(t)
\end{equation*}
So the resulting dynamic system is:
\begin{equation}
	\dot{\bm{x}}(t)=(\bm{A}-\bm{B}\bm{K})\bm{x}(t)+\bm{B}\bm{u}(t)
\end{equation}
\begin{equation}
	\bm{y}(t)=\bm{C}\bm{x}(t)
\end{equation}
And the resultant characteristic polynomial is:
\begin{equation}
    \det(s\bm{I}-\bm{A}+\bm{B}\bm{K})
\end{equation}
If the poles of this system are on the left side of the imaginary axis, then the system will be asymptotically stable.\\ The criterion we will use to select the QIP poles under this control will be by a second order system approximation, such approximation implies that the dominant poles can be calculated by:
\begin{equation}
\epsilon=\frac{|\ln{\frac{PO}{100}}|}{\sqrt{\pi^2+(\ln{\frac{PO}{100}})^2}}
\end{equation}
\begin{equation}
    \omega_n=\frac{4}{\epsilon t_s}
\end{equation}
\begin{equation}
    s_1=-\epsilon\omega_n+\omega_n\sqrt{\epsilon^2-1}
\end{equation}
\begin{equation}
    s_2=-\epsilon\omega_n-\omega_n\sqrt{\epsilon^2-1}
\end{equation}
It is important to emphasize that this second order approximation considers that the system is SISO. Considering an overshoot of $PO=1\%$ and settling time $t_s=6[s]$. Furthermore, under the criterion that the remaining 8 poles of the system will move away 10 times more than $s_1$ and $s_2$, following this procedure all poles of the system are determined. Having all poles, we can apply the pole placement method. Thus, the next feedback vector is obtained:
\begin{equation*}
	\scriptsize
	\bm{K}=
	\begin{bmatrix}
		11.01\\24.56\\-205.41\\-38.72\\-262.18\\-37.70\\147.26\\-35.33\\-1056.26\\-55.37
	\end{bmatrix}
\end{equation*}
Since this vector exists, QIP under this control is stable. To remove the steady-state error from this control we should use (\ref{error2}) obtaining that $N_{pp}=11.0072$.
\subsection{System simulation}
In order to test the stability and performance of the established control laws, the non linear dynamics of QIP will be simulated using Simscape of Matlab. As we mentioned earlier, this simulation environment is quite reliable when we want analyzing the performance of a dynamic system under control and take it to a functional prototyping.\\Images of the 3D view provided by Simscape and of the block diagrams designed for the simulation are attached below.
\begin{figure}[H]
	\centering
	\includegraphics[width=300px]{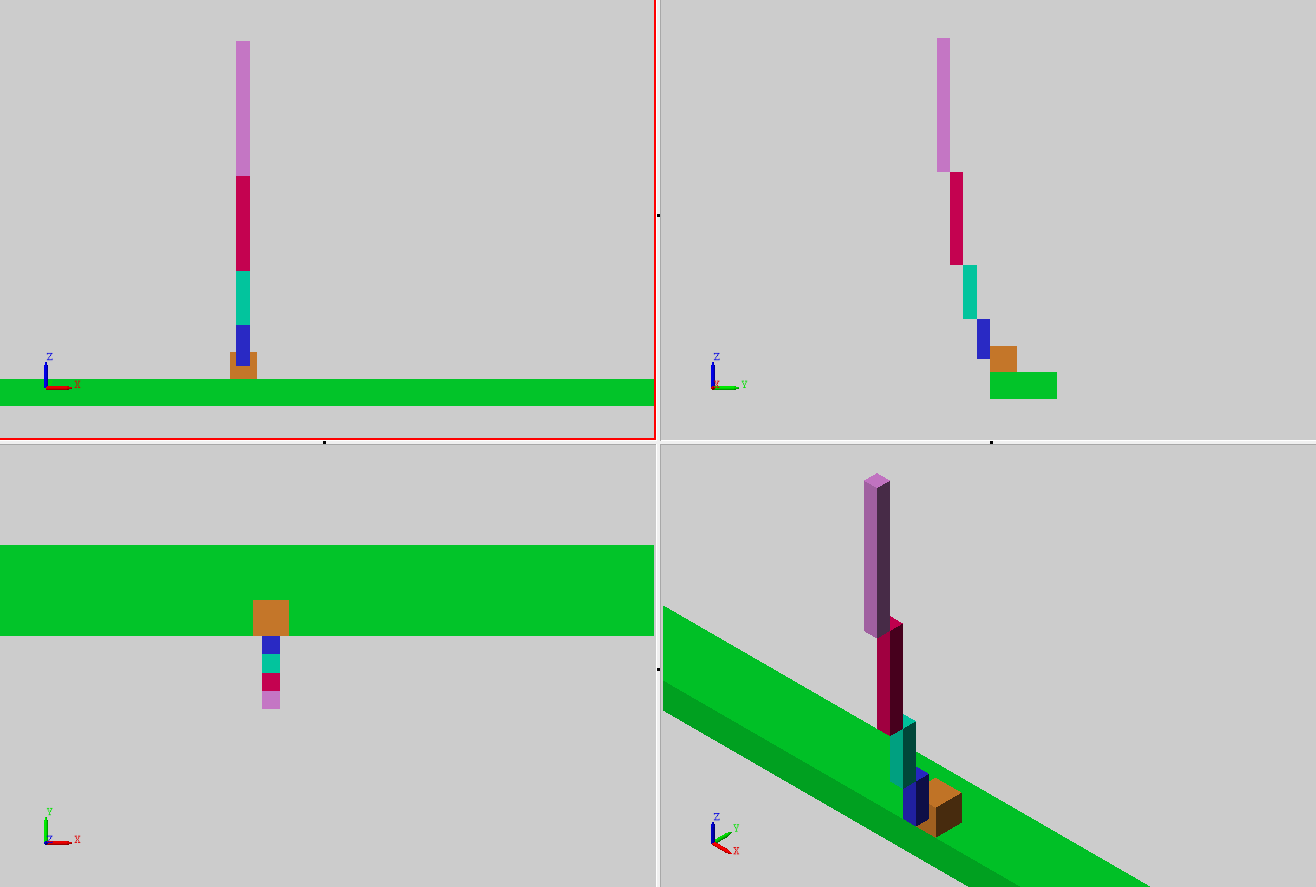}
	\caption{3D CAD view of QIP at Simscape}
\end{figure}
\begin{figure}[H]
	\centering
	\includegraphics[width=350px]{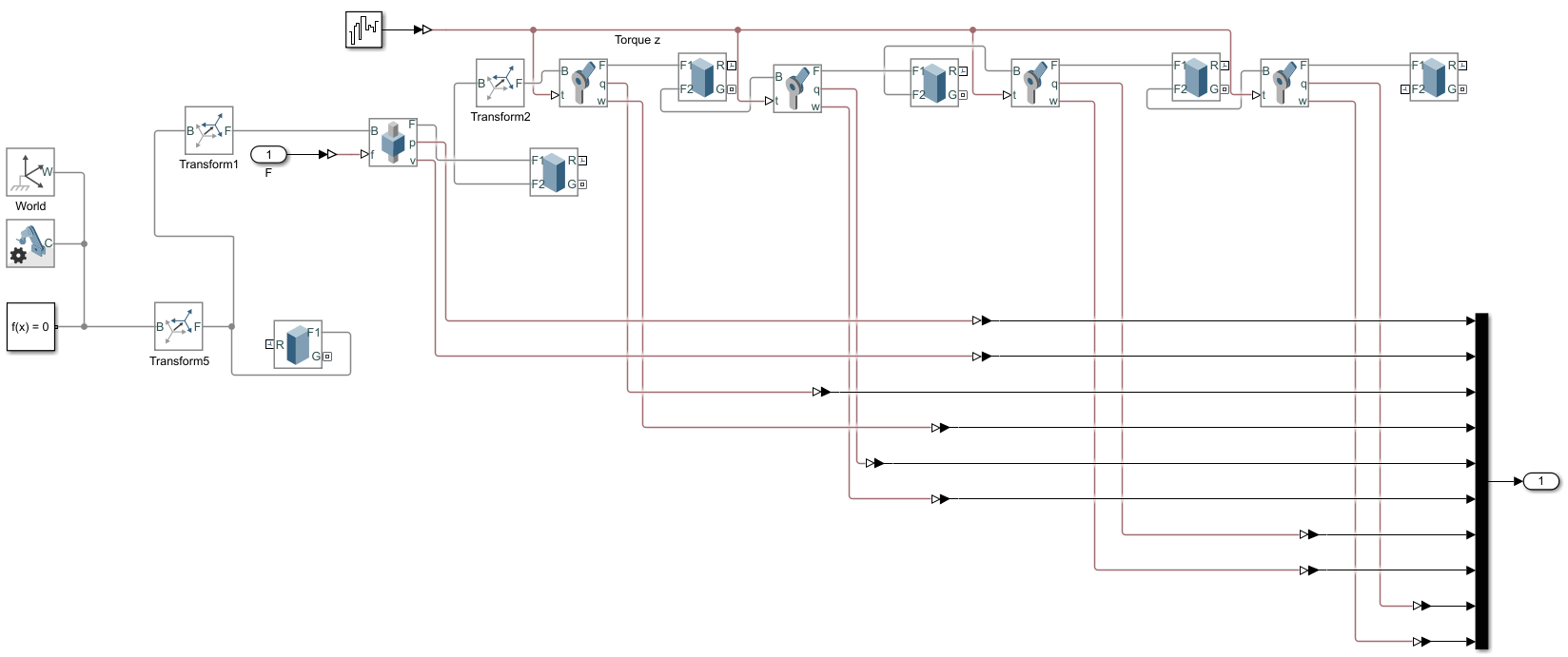}
	\caption{Block diagram of QIP, Simscape-Simulink}
\end{figure}
\begin{figure}[H]
	\centering
	\includegraphics[width=250px]{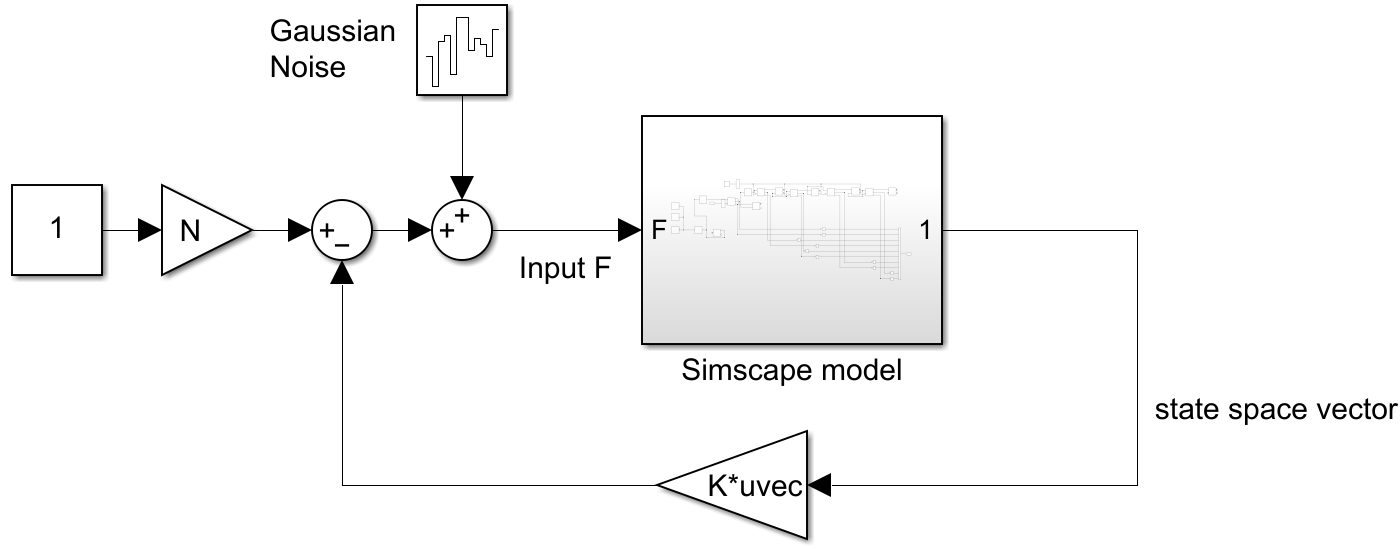}
	\caption{Close loop system, Simulink}
\end{figure}
\section{Results}
This section presents the responses of the system under control for both linear controllers in different conditions.\\
Responses to a reference step:
\begin{figure}[H]
	\centering
	\subfigure[Responses $x$ ]{\includegraphics[width=0.32\linewidth]{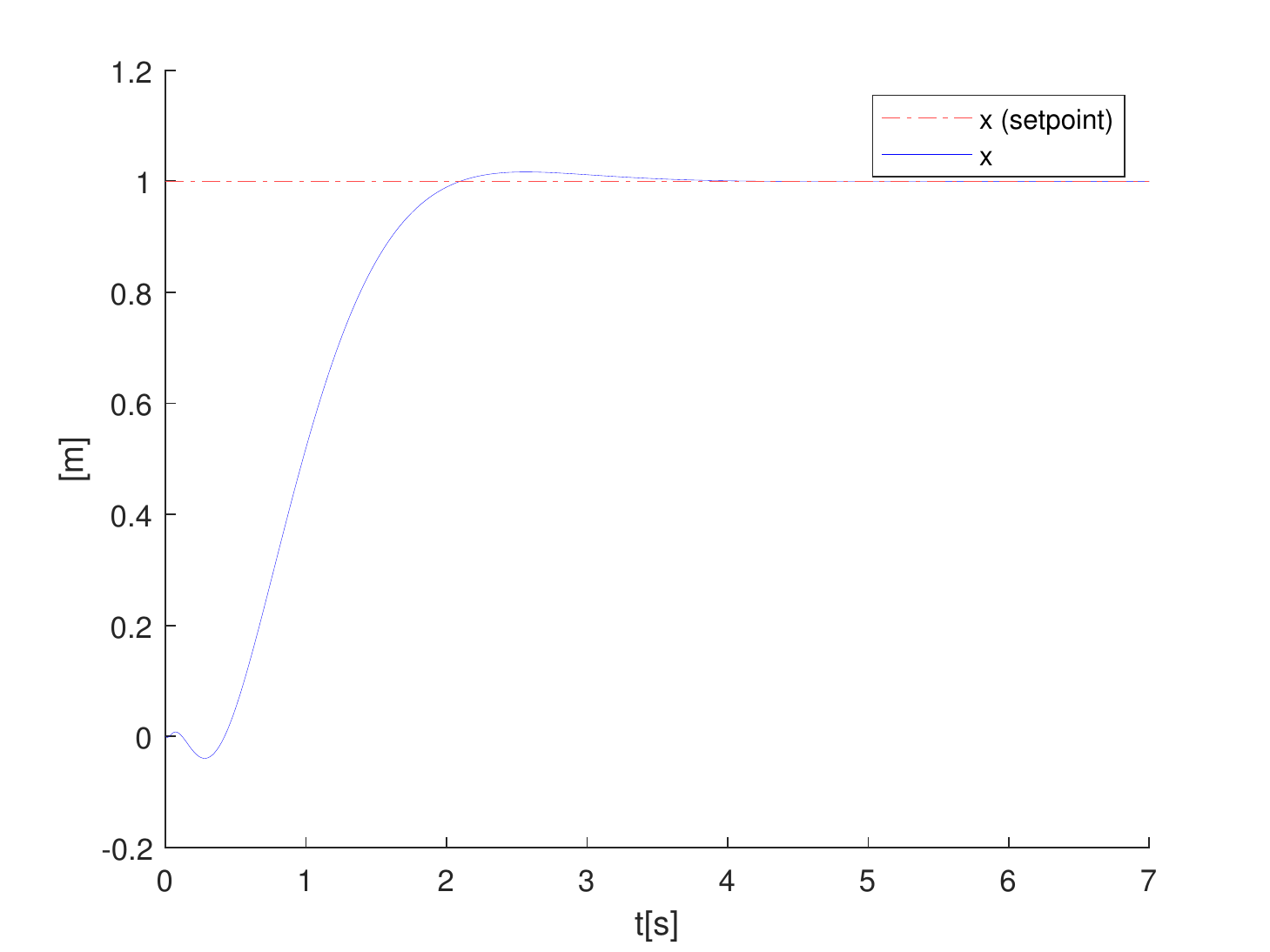}}
	\subfigure[Responses $\theta_i$]{\includegraphics[width=0.32\linewidth]{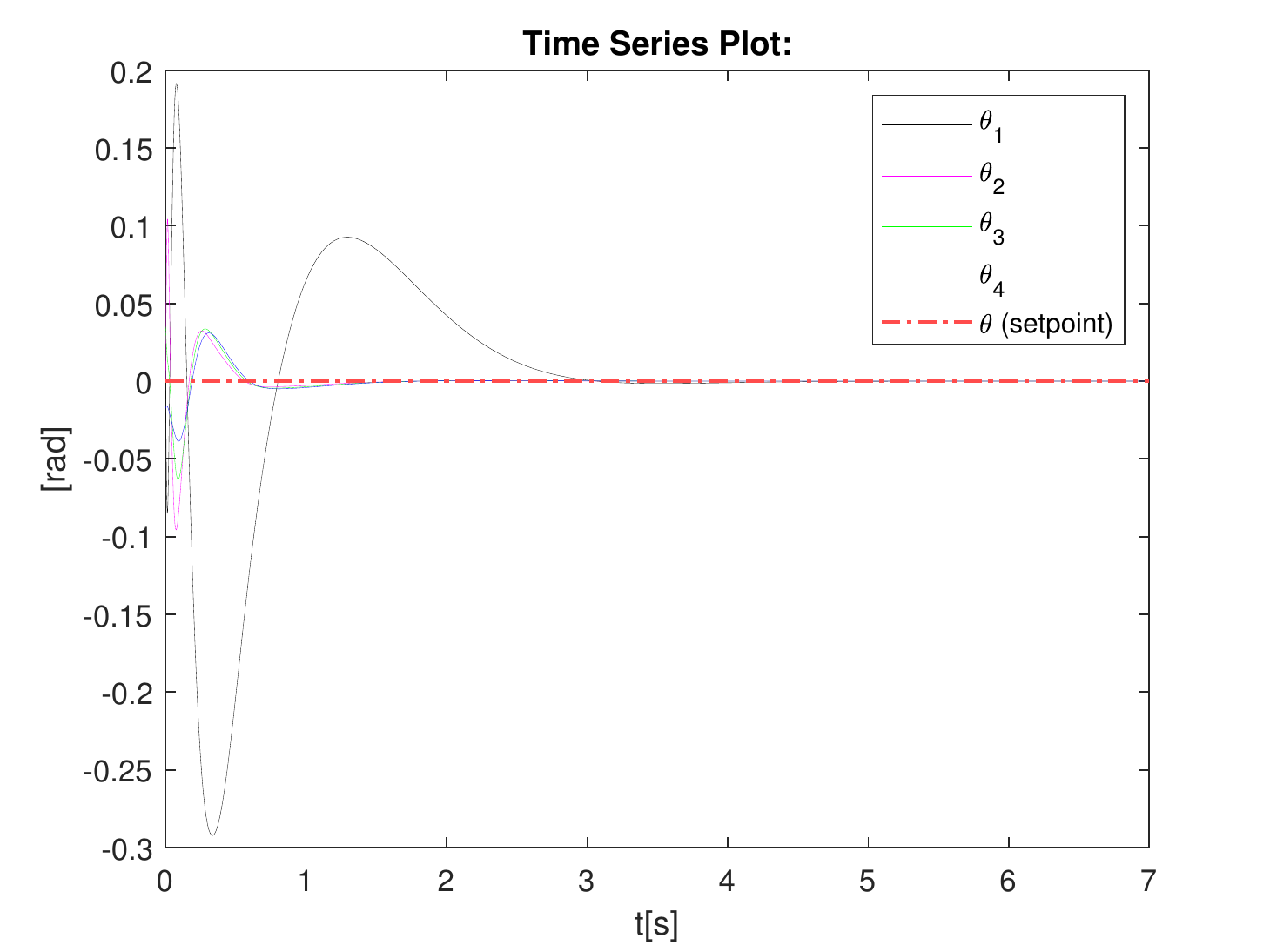}}
	\subfigure[Responses $\dot{\theta}_i$]{\includegraphics[width=0.32\linewidth]{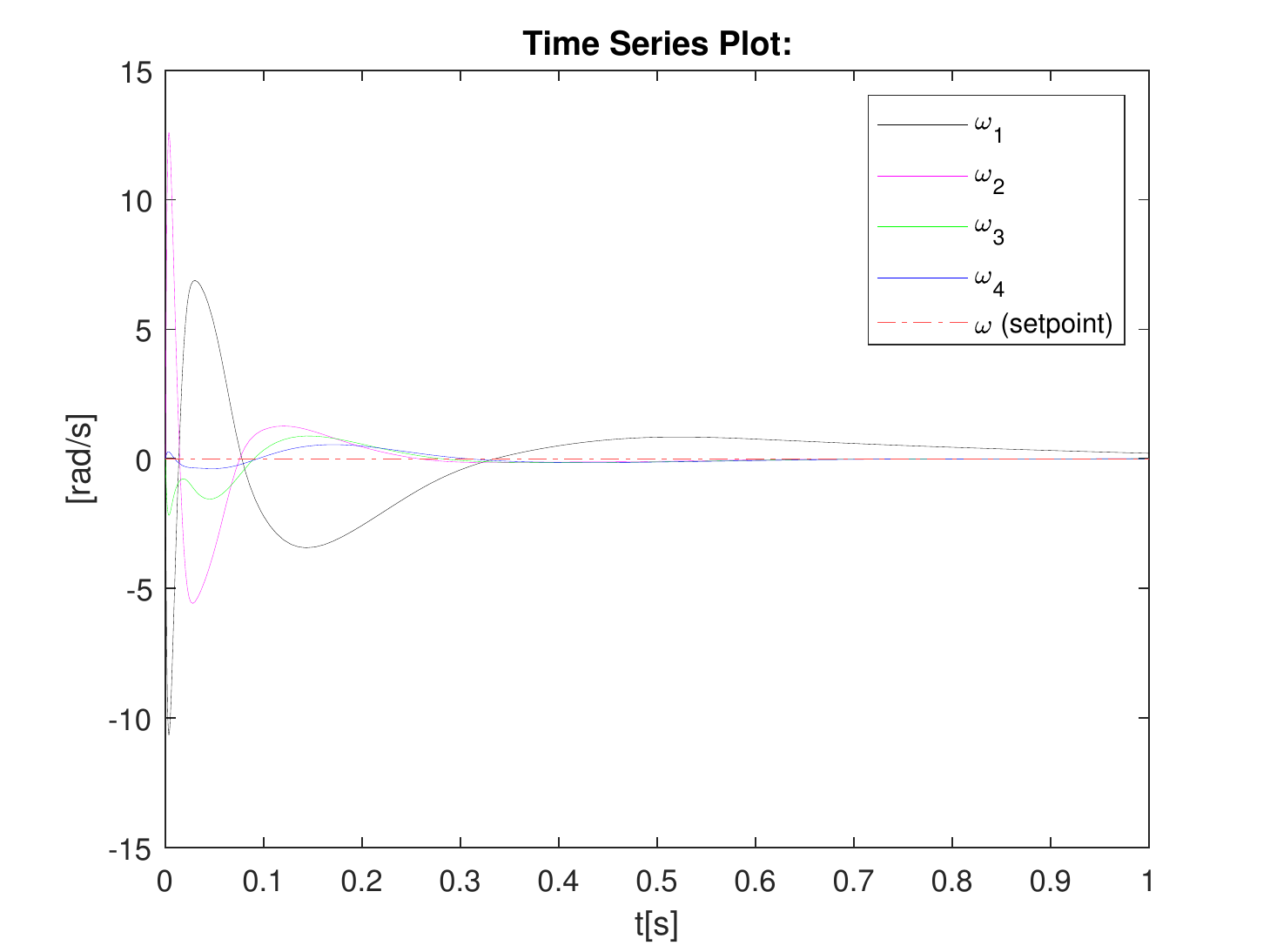}}
	\caption{Responses, through LQR control}
	\label{response_LQR}
\end{figure}
\begin{figure}[H]
	\centering
	\subfigure[Responses $x$ ]{\includegraphics[width=0.32\linewidth]{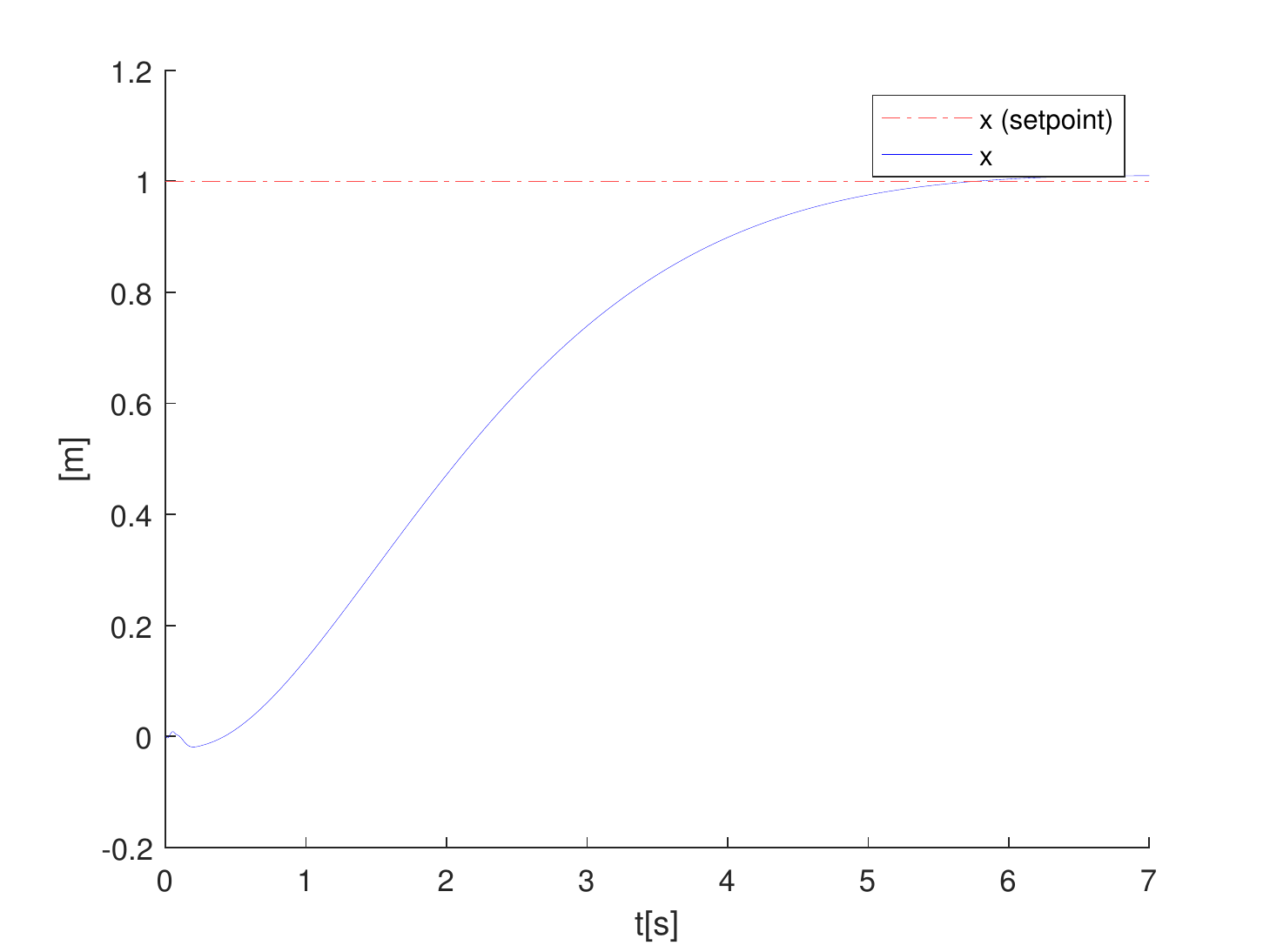}}
	\subfigure[Responses $\theta_i$]{\includegraphics[width=0.32\linewidth]{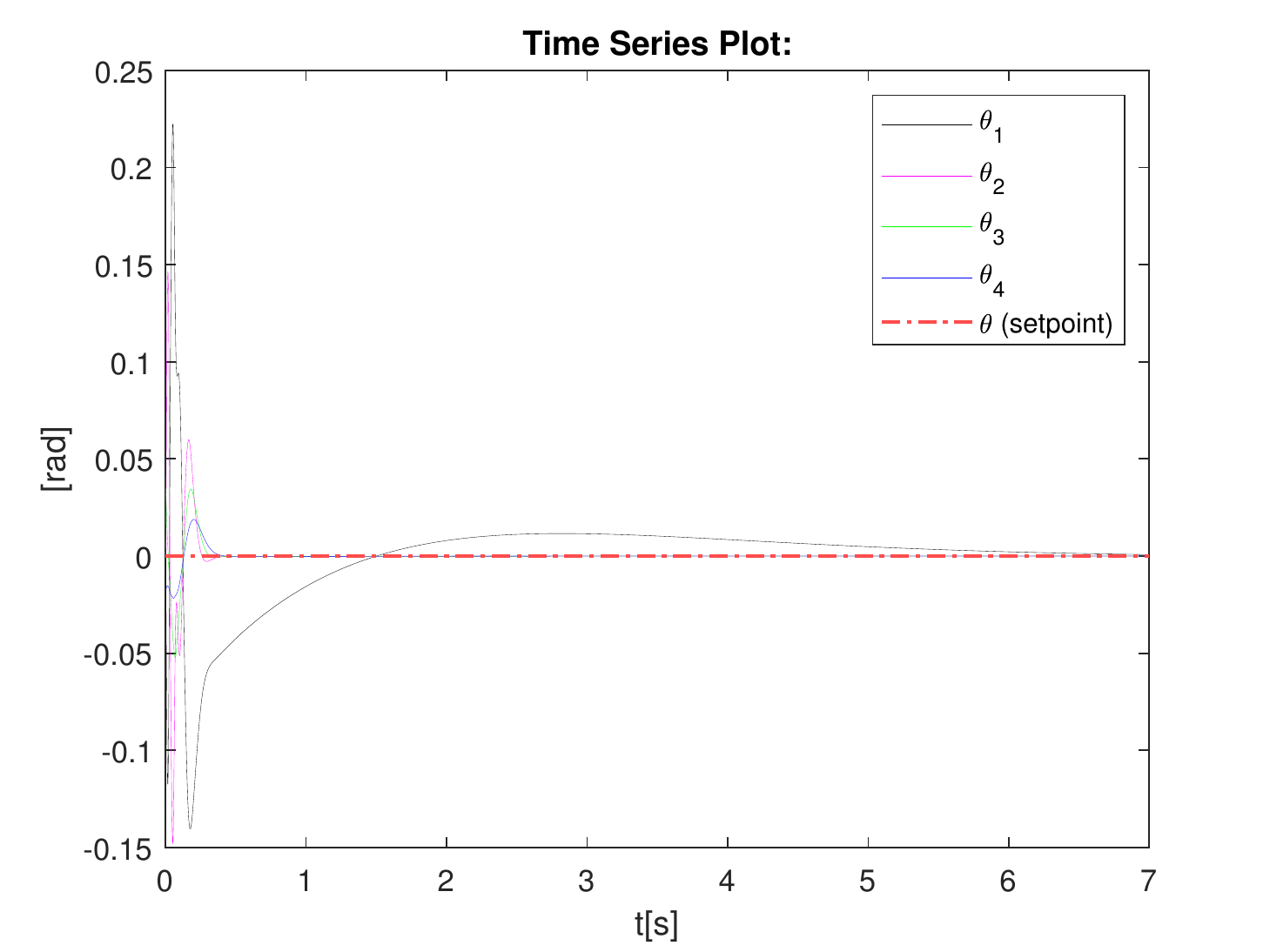}}
	\subfigure[Responses $\dot{\theta}_i$]{\includegraphics[width=0.32\linewidth]{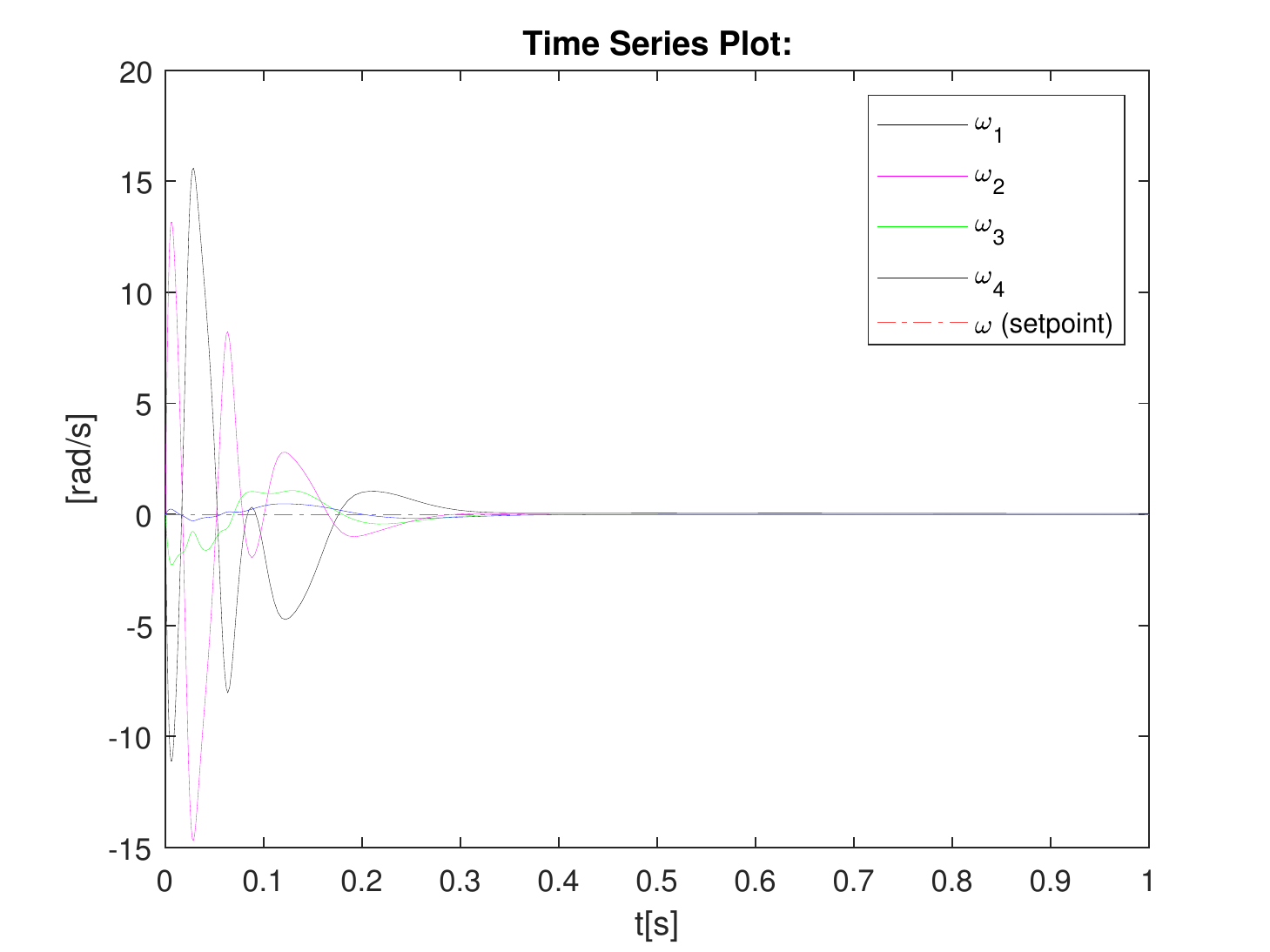}}
	\caption{Responses, through pole placement}
	\label{response_pole}
\end{figure}
Furthermore, we present a table with the results of the responses for pole placement technique control considering settling times of design, it is important emphasizing the feedback vector $\bm{K}$ was founded for every settling time.
\begin{table}[H]

\caption{Responses by pole placement for different settling time of design}
\label{table_2}
\centering

\begin{tabularx}{.6\linewidth}{XXX}
\toprule
Overshoot&Settling time& Stabilization \\ \midrule
	1\%&3[s]&No  \\ \midrule
	1\%&4[s]&No  \\ \midrule
	1\%&5[s]&Yes  \\ \midrule
	1\%&6[s]&Yes  \\ \midrule
	1\%&7[s]&Yes  \\ \midrule
	1\%&8[s]&No  \\ \midrule
	1\%&9[s]&No  \\ 
\bottomrule
\end{tabularx}
\end{table}
Now, we present the response of the system under both controllers designed, disturbance input $F_d\sim\mathcal{N}(0,0.01)$ and disturbance torque in each link $(\tau_d)_i\sim\mathcal{N}(0,10^{-9})$ with $i\in\{1,2,3,4\}$.
\begin{figure}[H]
	\centering
	\subfigure[Responses $x$ ]{\includegraphics[width=0.32\linewidth]{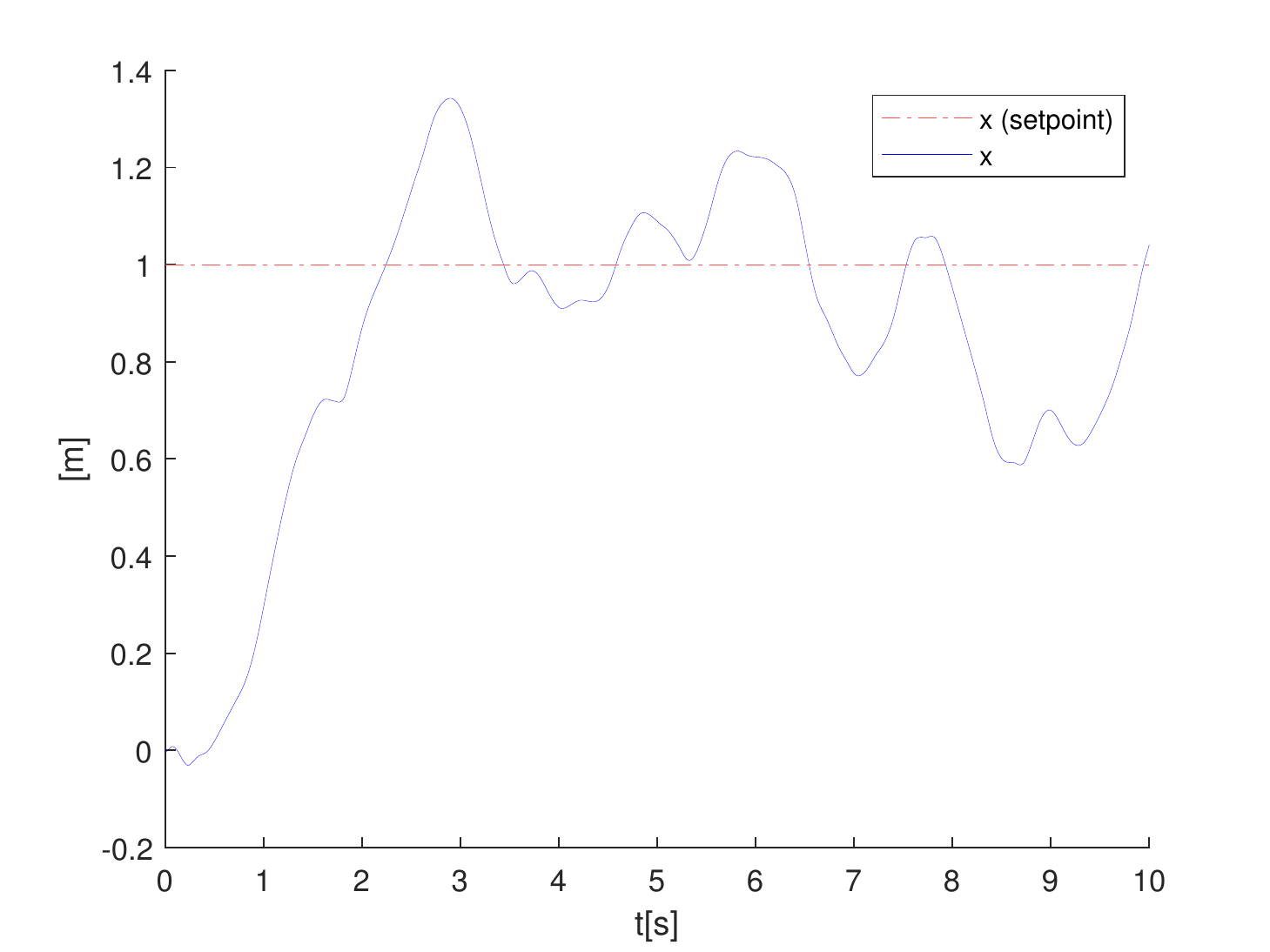}}
	\subfigure[Responses $\theta_i$]{\includegraphics[width=0.32\linewidth]{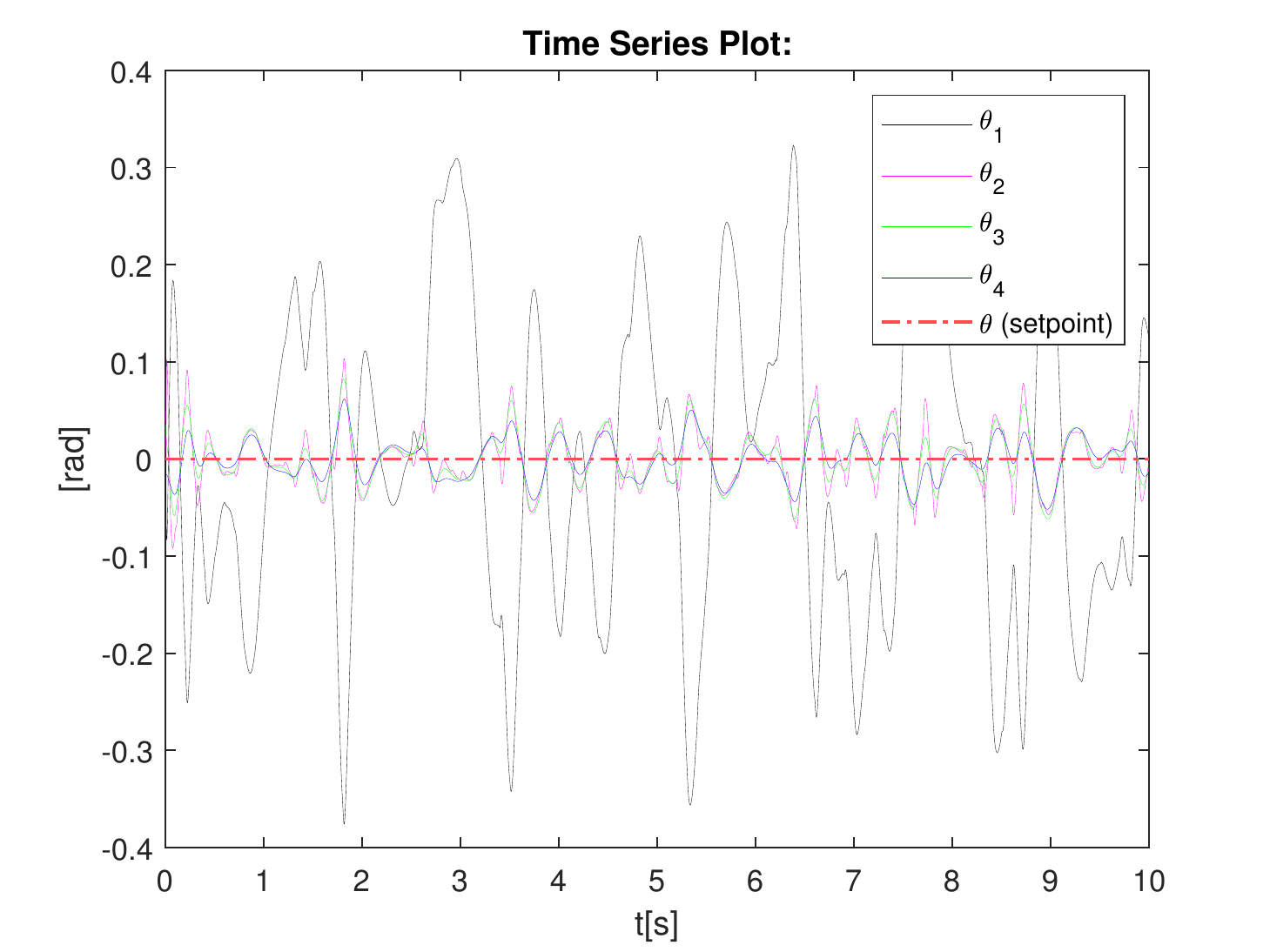}}
	\subfigure[Responses $\dot{\theta}_i$]{\includegraphics[width=0.32\linewidth]{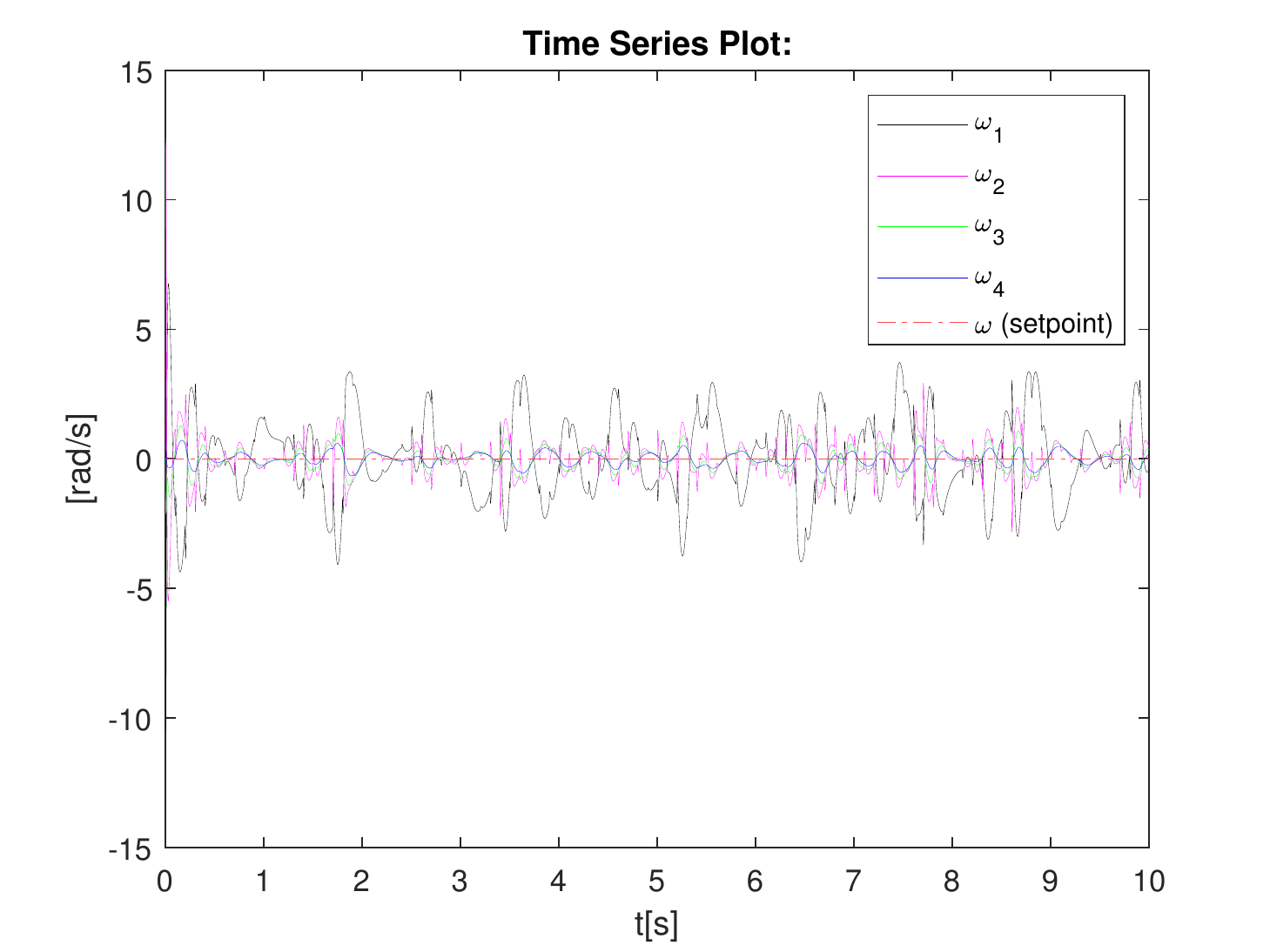}}
	\caption{Responses, under LQR control and disturbances}
	\label{response_LQR_dist}
\end{figure}
\begin{figure}[H]
	\centering
	\subfigure[Responses $x$ ]{\includegraphics[width=0.32\linewidth]{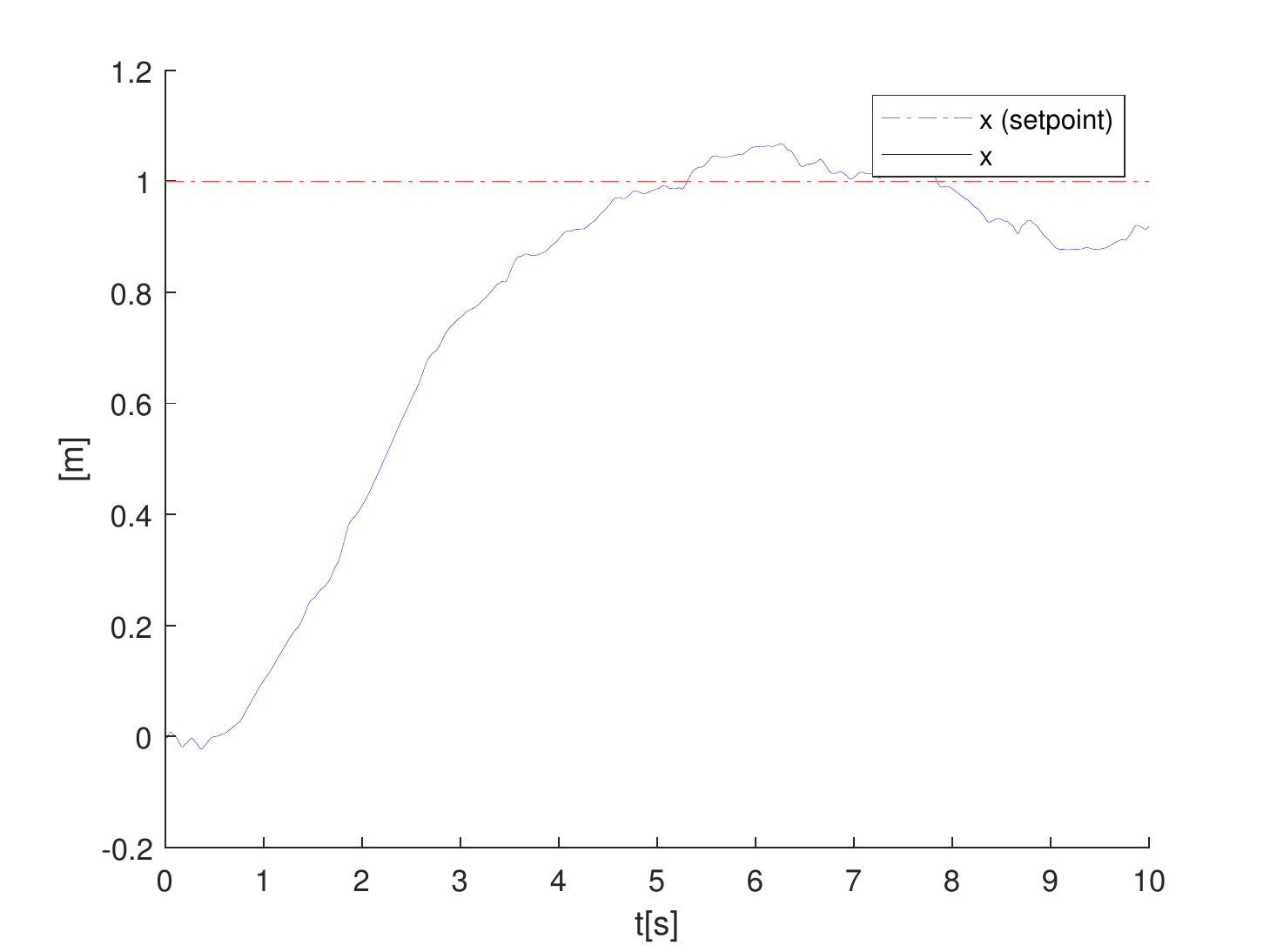}}
	\subfigure[Responses $\theta_i$]{\includegraphics[width=0.32\linewidth]{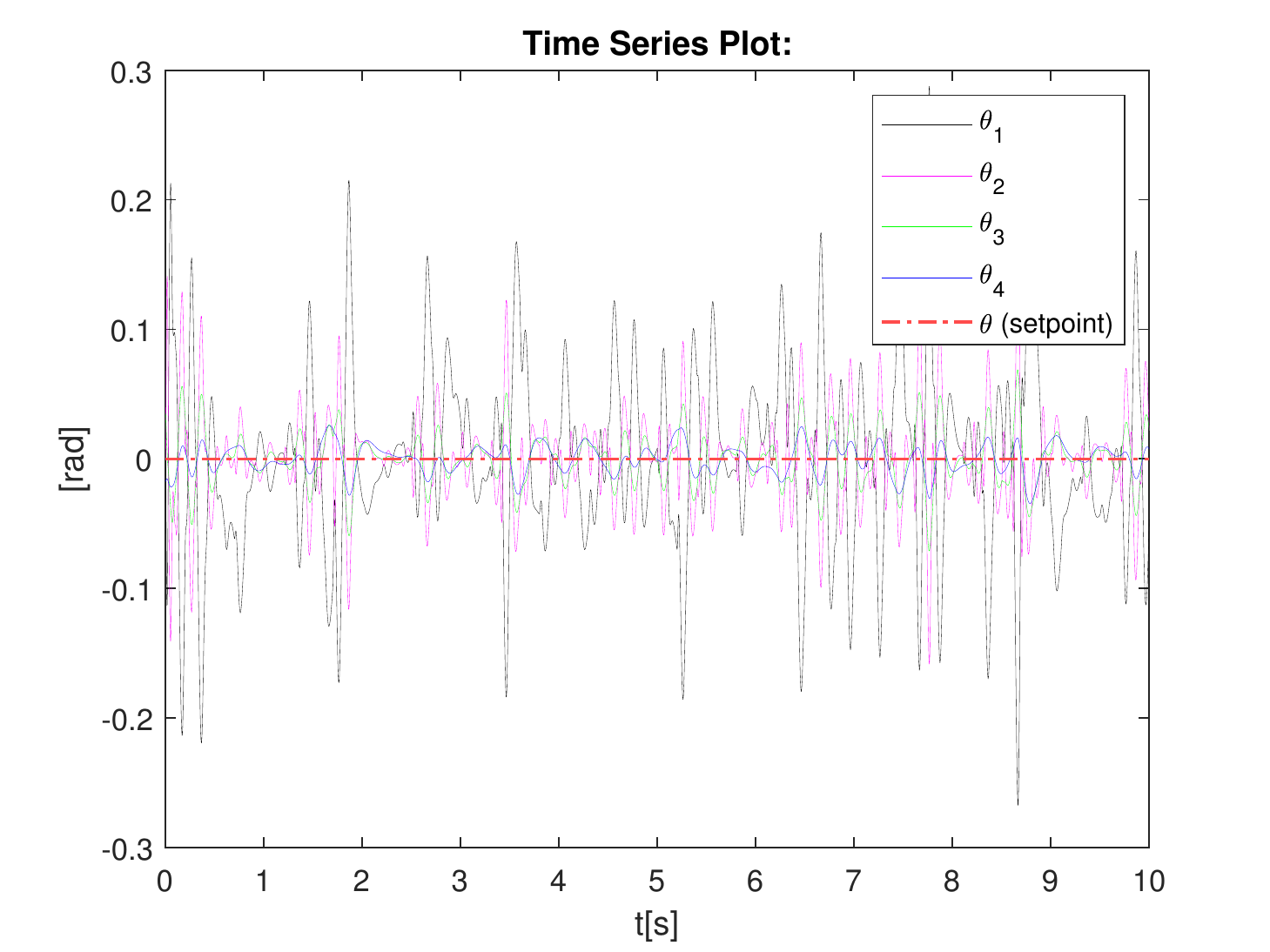}}
	\subfigure[Responses $\dot{\theta}_i$]{\includegraphics[width=0.32\linewidth]{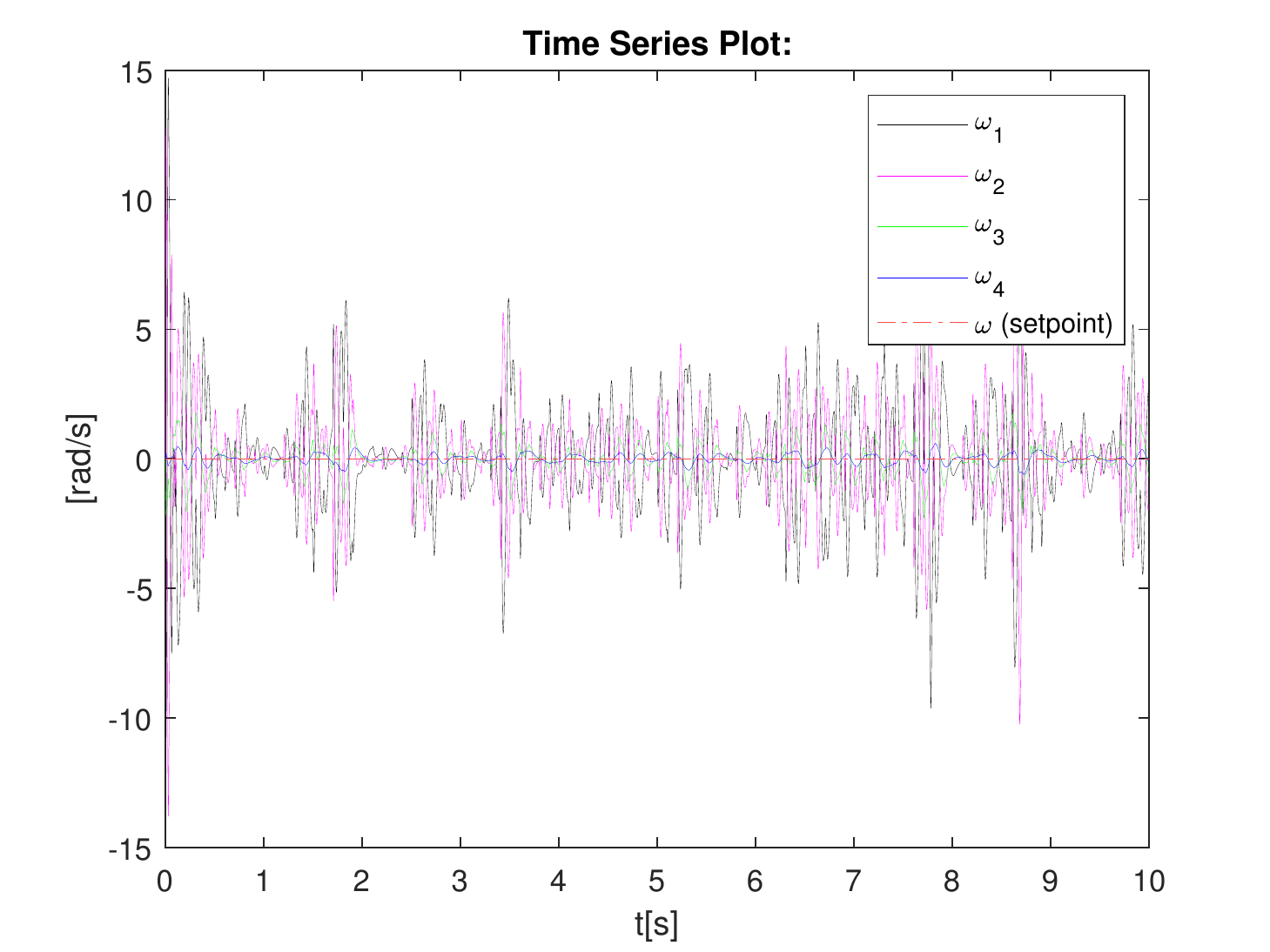}}
	\caption{Responses, under pole placement technique control and disturbances}
	\label{response_pole_dist}
\end{figure}
In addition, these results can be seen in the following video:
\begin{center}
\textcolor{blue}{\underline{\url{https://youtu.be/hjg5mrtI8nA}}}    
\end{center}

\section{Discussion}
The results presented in the figure \ref{response_LQR} and \ref{response_pole} indicate that both LQR control and pole placement technique stabilize the plant at the established set-point. The results obtained with pole placement technique indicate that the input requirements (settling time and overshoot) are reach because of the desired settling time and overshoot of the system under control is 
approximately equal to the proposed. However, as we can see in the table \ref{table_2}, these parameters of design are limited due to in some cases the stabilization is not reach due to the second order approximation for designing this control only consider the position of the cart, thus, we don't have a criteria for avoiding the other states get out of them linear zone.\\
On the other hand, the figure \ref{response_LQR} shows that the response of the system under LQR also stabilizes the plant to the desired set-point. Even though, we don't have a criteria for the overshoot and settling time LQR is very effective when only the stabilization of the plant is concerned. It is due to this control generates a feedback vector that can give more or less energy importance to the states of the plant and control input depending on the weight matrices $\bm{Q}$ and $\bm{R}$. Thus, we have a way to quantify that the control law does not go out of its linear zone. Besides the figure \ref{response_LQR_dist} and \ref{response_pole_dist} show that the response of the plant under LQR control has less frequency than response of pole placement technique control. Thus, LQR control is more robust.
\section{Conclusion}
This paper has presented a workflow for obtaining the mathematical model of an inverted pendulum with $n$ links mounted on a car where the angles are measured with respect to the previous link. The case $n=4$ has studied in this work. Furthermore, we have presented a workflow for designing two linear controllers LQR and pole placement technique with a pre compensation gain for eliminating the steady state error and we validated these results doing an non linear simulation using Simscape of Matlab. The results indicate that although we can obtain a feedback vector that produces an specific setting time and overshoot in the position of the cart through pole placement technique and second order approximation approach, these two parameters of design are very limited and also produce a response with more frequency than LQR control. On the other hand, LQR approach allowed us to have a mathematical criteria for avoiding that all states of QIP outside of linearization zone, it based on the value of weight matrices $\bm{Q}$ and $\bm{R}$ and thus avoid instability.\\

\bibliographystyle{unsrt} 
\bibliography{references}

\begin{thebibliography}{10}

\bibitem{bradshaw_shao_1996}
Alan Bradshaw and Jindi Shao.
\newblock Swing-up control of inverted pendulum systems.
\newblock {\em Robotica}, 14(4):397–405, 1996.

\bibitem{zurawska_reconfigurable_2017}
Magdalena Żurawska, Maksymilian Szumowski, and Teresa Zielińska.
\newblock Reconfigurable {Double} {Inverted} {Pendulum} {Applied} to the
  {Modelling} of {Human} {Robot} {Motion}.
\newblock {\em Journal of Automation, Mobile Robotics \& Intelligent Systems},
  11(2):12--20, June 2017.

\bibitem{zhen_inverted_2018}
Bin Zhen, Liang Chang, and Zigen Song.
\newblock An {Inverted} {Pendulum} {Model} {Describing} the {Lateral}
  {Pedestrian}-{Footbridge} {Interaction}.
\newblock {\em Advances in Civil Engineering}, 2018:1--12, November 2018.

\bibitem{collini_oscillations_2016}
Luca Collini, Rinaldo Garziera, Kseniia Riabova, Mariya Munitsyna, and
  Alessandro Tasora.
\newblock Oscillations {Control} of {Rocking}-{Block}-{Type} {Buildings} by the
  {Addition} of a {Tuned} {Pendulum}.
\newblock {\em Shock and Vibration}, 2016:1--11, 2016.

\bibitem{higher_order}
Saibal Manna, Shaili Shaw, and Ashok~Kumar Akella.
\newblock Design of higher-order regulator system using pole-placement
  technique.
\newblock In {\em 2020 International Conference on Computational Intelligence
  for Smart Power System and Sustainable Energy (CISPSSE)}, pages 1--4, 2020.

\bibitem{PID_pole}
Furkan~Nur Deniz, Baris~Baykant Alagoz, and Nusret Tan.
\newblock Pid controller design based on second order model approximation by
  using stability boundary locus fitting.
\newblock In {\em 2015 9th International Conference on Electrical and
  Electronics Engineering (ELECO)}, pages 827--831, 2015.

\bibitem{8619302}
X.~{Xin}, K.~{Zhang}, and H.~{Wei}.
\newblock Linear strong structural controllability for an n-link inverted
  pendulum in a cart.
\newblock In {\em 2018 IEEE Conference on Decision and Control (CDC)}, pages
  1204--1209, 2018.

\bibitem{7976186}
J.~{Königsmarková} and M.~{Schlegel}.
\newblock Identification of n-link inverted pendulum on a cart.
\newblock In {\em 2017 21st International Conference on Process Control (PC)},
  pages 42--47, 2017.

\bibitem{LQR1}
Tamen~Thapa Sarkar and Lillie Dewan.
\newblock Application of lqr and mrac for swing up control of inverted
  pendulum.
\newblock In {\em 2017 4th International Conference on Power, Control Embedded
  Systems (ICPCES)}, pages 1--6, 2017.

\bibitem{LQR2}
Xin Xiong and Zhou Wan.
\newblock The simulation of double inverted pendulum control based on particle
  swarm optimization lqr algorithm.
\newblock In {\em 2010 IEEE International Conference on Software Engineering
  and Service Sciences}, pages 253--256, 2010.

\bibitem{simscape1}
Omer Eldirdiry and Riadh Zaier.
\newblock Modeling biomechanical legs with toe-joint using simscape.
\newblock In {\em 2018 11th International Symposium on Mechatronics and its
  Applications (ISMA)}, pages 1--7, 2018.

\bibitem{simscape2}
Hubert Milanowski and Adam~Krzysztof Piłat.
\newblock Comparison of identified and simscape model of human leg motion.
\newblock In {\em 2020 International Conference Mechatronic Systems and
  Materials (MSM)}, pages 1--6, 2020.

\bibitem{simscape3}
Kichang Lee, Jiyoung Lee, Bungchul Woo, Jeongwook Lee, Young-Jin Lee, and
  Syungkwon Ra.
\newblock Modeling and control of a articulated robot arm with embedded joint
  actuators.
\newblock In {\em 2018 International Conference on Information and
  Communication Technology Robotics (ICT-ROBOT)}, pages 1--4, 2018.

\bibitem{simscape4}
Shuvra Das.
\newblock {\em Modeling and Simulation of Mechatronic Systems using Simscape}.
\newblock 2020.

\bibitem{973983}
{Wei Zhong} and H.~{Rock}.
\newblock Energy and passivity based control of the double inverted pendulum on
  a cart.
\newblock In {\em Proceedings of the 2001 IEEE International Conference on
  Control Applications (CCA'01) (Cat. No.01CH37204)}, pages 896--901, 2001.

\end{thebibliography}

\end{document}